\documentclass[prx,showpacs,twocolumn,superscriptaddress]{revtex4-1}
\usepackage{graphicx,amsmath,amssymb,dcolumn,subfigure}
\usepackage{epsfig}
\usepackage[normalem]{ulem}  

\newcommand{\av}[1]{\left\langle #1 \right \rangle}
\newcommand{\vc}[1]{{\bf #1}}

\begin{document}
\title{Trains, tails and loops of partially adsorbed semi-flexible filaments}
\author{David Welch}
\affiliation{Graduate Program in Biophysics and Structural Biology, Brandeis University,
 Waltham, MA 02454, USA}
\author{M. P. Lettinga}
\affiliation{Soft Matter, Institute of Complex Systems, Forschungszentrum J\"ulich, D-52425 J\"ulich, Germany. KU Leuven, Laboratory for Soft Matter and Biophysics, , Celestijnenlaan 200D, B-3001 Leuven, Belgium.}
\author{Marisol Ripoll}
\affiliation{Theoretical Soft Matter and Biophysics, Institute of Complex Systems, Forschungszentrum J\"ulich, D-52425 J\"ulich, Germany}
\author{Zvonimir Dogic} 
\email{zdogic@brandeis.edu} 
\affiliation{Department of Physics, Brandeis University, Waltham, MA 02454, USA}
\author{Gerard A. Vliegenthart}
\email{g.vliegenthart@fz-juelich.de} 
\affiliation{Theoretical Soft Matter and Biophysics, Institute for Advanced Simulation, Forschungszentrum J\"ulich, D-52425 J\"ulich, Germany}
\date{\today}

\begin{abstract}
Polymer adsorption is a fundamental problem in statistical mechanics that has direct relevance to diverse disciplines ranging from biological lubrication to stability of colloidal suspensions. We combine experiments with computer simulations to investigate depletion induced adsorption of semi-flexible polymers onto a hard-wall. Three dimensional filament configurations of partially adsorbed F-actin polymers are visualized with total internal reflection fluorescence microscopy. This information is used to determine the location of the adsorption/desorption transition and extract the statistics of trains, tails and loops of partially adsorbed filament configurations. In contrast to long flexible filaments which primarily desorb by the formation of loops, the desorption of stiff, finite-sized filaments is largely driven by fluctuating filament tails. Simulations quantitatively reproduce our experimental data and allow us to extract universal laws that explain scaling of the adsorption-desorption transition with relevant microscopic parameters. Our results demonstrate how the adhesion strength, filament stiffness, length, as well as the configurational space accessible to the desorbed filament can be used to design the characteristics of filament adsorption and thus engineer properties of composite biopolymeric materials.
\end{abstract}

\maketitle
\section{Introduction}
In comparison to simple rigid molecules whose adsorption can be described by straightforward two-state models, adsorption of complex molecules with internal degrees of freedom, such as polymeric chains, require development of more intricate theories. The complication arises because certain segments of an adsorbed filament can be surface bound while other segments within the same molecule remain desorbed. This qualitative change in behavior requires development of theoretical models of adsorption transition that can account for internal degrees of freedom of polymeric chains~\cite{SKVORTSOV76,BIRSHTEIN79,MAGGS89,SEMENOV96, KUZNETSOV97,NETZ99, BOEHM06, TANG010, LINSE010,KLUSHIN013,NETZ,KRAMARENKO96}. At a fundamental level the configurations of a partially adsorbed polymer can be described in terms of trains, tails and loops~\cite{vanderLINDEN96,SILBERBERG62,HOEVE65,SEMENOV96}. Trains are polymer segments that are adsorbed onto a surface, loops are desorbed segments separated by two trains while tails account for the desorbed end segments. Traditional scattering-based experimental techniques used to investigate polymer adsorption do not directly reveal the statistics of tails, trains and loops~\cite{COHEN-STUART,AUVRAY}. Instead they yield structural information that is averaged over many states and molecules and as as such it cannot be used to directly extract conformational statistics of partially adsorbed filaments.

Microscopy has significantly advanced polymer science by enabling single-molecule visualization of structures and dynamical processes
that could previously be only studied by bulk techniques. It has been used to visualize dynamics of desorbed 3D polymers as well as
filaments that are strongly adsorbed onto the substrate so that they effectively behave as a 2D system~\cite{PERKINS1,PERKINS2,KAS,MAIER,RIVETTI}. Here we bridge these two previously studied limits by visualizing 3D configurations of semi-flexible filaments as they undergo an adsorption/desorption transition. We examine how the strength of the polymer-wall attractions affects the statistics of trains, tails and loops. This information is used to estimate the location of
the adsorption-desorption transition. We study the same processes using computer simulations and find quantitative agreement with the experimental measurements. In contrast to desorption of flexible chains which is dominated by loop formation, the desorption of finite length semi-flexible filaments is largely driven by tail fluctuations. With decreasing attraction strength the average tail length increases up to the point where it is comparable to the filament contour length, whereupon the entire filament desorbs from the surface. The statistics of filament loops, tails and trains and their dependence on attraction strength is in broad agreement with computer simulation that have previously studied similar phenomena~\cite{HSU013,SANCHEZ011}

\section{Experimental methods}
\begin{figure}[h!]
\includegraphics[width=0.45\textwidth]{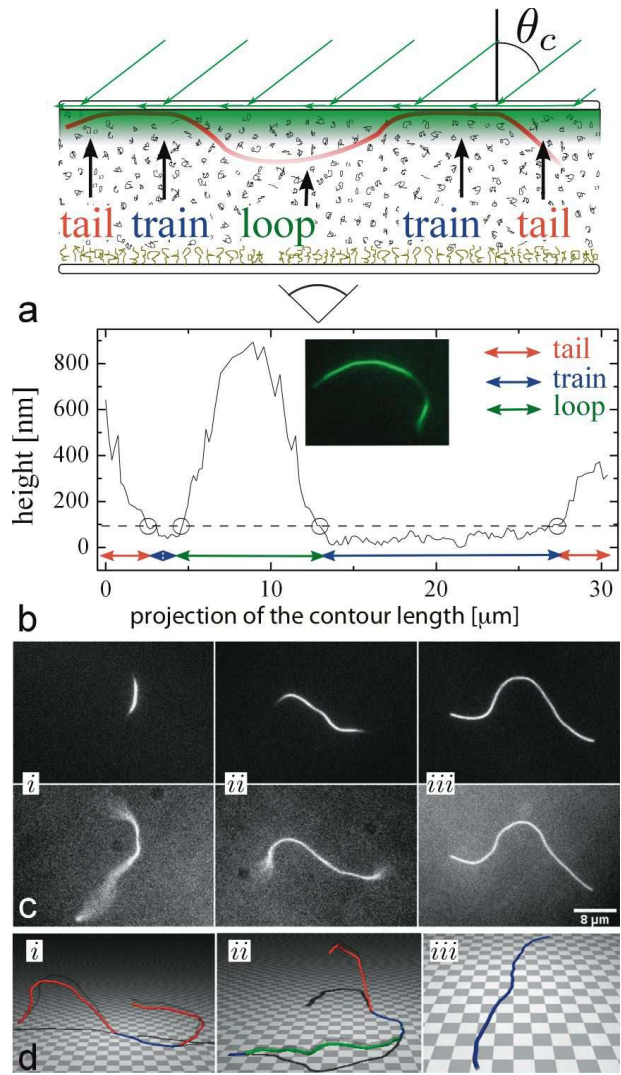}
\caption{TIRF microscopy determines 3D conformations of a fluctuating
  filament undergoing adsorption/desorption transition. {\textbf a)}
  Schematic of the experimental setup. An evanescent field is achieved
  by refracting laser through a prism above the critical angle
  $\theta_c$. The sample is viewed from below with a high numerical aperture (NA)
  objective. A polymer brush coating the coverslip surface ensures
  that filaments adsorb only to the top surface.  {\textbf b}) The
  filament height plotted as a function of its position makes it
  possible to identify tails, trains and loops. The local height, $h$,
  is extracted from the filament intensity according to the
  relationship, $h=-a\ln(I/I_0)$ where $I$ is the intensity at a given
  height, $I_0$ the contour's maximum intensity (an offset is
  subtracted from both) and $a$ is a decay constant of the evanescent
  field. Inset: Image of the filaments from which the height
  information is extracted. 
  {\textbf c}) TIRF (top) and corresponding epi-fluorescence (bottom) images of
  adsorbed filaments at three different dextran concentrations
  corresponding to (i) the desorbed regime, (ii) the transition region
  and (iii) strongly adsorbed regime. For related movies see Ref.~\cite{movie}.  %
  {\bf d}) Snapshots of filament configuration obtained with
    computer simulations corresponding to three regime.
}\label{fig:definitions}
\end{figure}

For experimental work we used phalloidin stabilized F-actin filaments,
which have 6 nm diameter and a persistence length of ~16
$\mu$m~\cite{HOWARD}. The large persistence length allowed us to
reconstruct the entire 3D filament conformation. Negatively charged
F-actin filaments are repelled from a glass surface of the same
charge. To induce a tunable filament-wall attraction we added
non-adsorbing polymer dextran (MW 500,000, Sigma), which has a radius
of gyration $R_g \sim$ 18 nm~\cite{SENTI}. The dextran center of mass
is excluded from a layer of thickness $R_g$ adjacent to a hard wall as
well as from a shell of radius $R_g+R_f$, where $R_f=3$nm is the radius of the filament, surrounding each filament. As
F-actin approaches the wall, the two excluded volume regions overlap,
leading to increase in the total volume accessible to the depleting
polymers and attractive wall-filament depletion
interactions~\cite{ASAKURA,DINSMORE}. All experiments were performed
in the dilute regime, ensuring that the
actin-substrate interaction strength is linearly proportional to the dextran
concentration. The instantaneous 3D configurations of a partially adsorbed filament
were extracted from Total Internal Reflection Fluorescence (TIRF)
Microscopy images. In the evanescent field the excitation intensity
decays exponentially; consequently, the filament distance away from
the wall is directly related to its local
intensity~\cite{AXELROD}. 
Combining depletion interaction with TIRF
imaging allowed us to tune the strength of filament-wall attractive
potential while simultaneously visualizing the entire 3D filaments
configuration with nanometer accuracy (Fig.\ref{fig:definitions}a, b).

For reproducible results we used cover slides from the same production
batch (VWR). The slides where incubated in 6 M HCl for 45 minutes,
rinsed with deionized water and sonicated in hot soap water (1\%
Hellmanex, Hellma) for five minutes before final rinsing and storage
in ddH$_2$O. Slides were used within 24 hours of cleaning. To adsorb
filaments only on one surface, the coverslip surface was coated with a
poly-acrylamide brush which suppresses wall-filament depletion
interactions~\cite{LAU,SANCHEZ13}. Actin was column-purified, polymerized and
labelled with Alexa-488 phalloidin, at 1:1 monomer:dye
ratio~\cite{PARDEE}. F-actin filaments were dissolved in buffer
containing 20 mM NaH$_2$PO$_4$ (pH=7.4), 200 mM KCl, and 30\% sucrose
(w/v). Sucrose slows down desorption of Alexa-488 phalloidin and thus
prolongs the observation time of F-actin filaments~\cite{CRUZ}. Actin
filaments are inherently polydisperse; for our analysis we only used
filaments that are larger than the persistence length. The average
length of the analyzed filaments was~33 $\mu$m which is roughly twice
the persistence length. For each dextran concentration we have
analyzed on average five different filaments of comparable
contour length.

We used a prism-based TIRF setup since it produces evanescent fields
with low background fluorescence and a better defined exponential
decay length~\cite{AXELROD}. Phalloidin labelled actin filaments
exhibit a pronounced polarization dependent fluorescence
signal~\cite{DISCHER}. In order to minimize this contribution we
ensured that the polarization of the laser was in the plane defined by
the incident and reflected light~\cite{AXELROD}. To determine the
accuracy of our method we have imaged filaments that are irreversibly
attached to the surface and thus exhibited no height
fluctuations. Intensity fluctuation of such filaments corresponded to
apparent height variations of $\pm$40 nm; hence we estimate that our
resolution is about 80 nm. The fluorescence images were acquired using
an EMCCD operating in a conventional mode (Andor iXon), mounted on a
Nikon ecclipseTE-2000 microscope and using a PlanFluor-100x NA
  1.3 objective. The time between exposures was 2 seconds ensuring
that the subsequent images were uncorrelated. We imaged a single
filament for approximately 200 frames before photobleaching effects
began to influence chain statistics. In parallel with TIRF imaging,
filaments were simultaneously visualized with epi-fluorescence
microscopy which enabled determination of the filament contour
length. To experimentally measure the evanescent field decay constant
we prepared chambers of well-defined thickness that ranged between 1
and 2 $\mu$m with 100 nm beads adsorbed onto both surfaces. The
chamber thickness was accurately measured by an objective mounted on a
piezo-stage. Knowing the chamber thickness and measuring the ratio of
bead intensities attached to opposite surface directly yields the
decay constant of the evanescent field.  The TIRF decay length
measured in this way was $\Xi=230\pm40$ nm. \textbf{This is close to theoretical
predictions $\Xi=\frac{\lambda}{2\pi(\sin\theta-\sin\theta_c)^{1/2}}$, where $\theta_c$ is the critical angle,  $\theta = 61°$ is the incident angle away from the normal to the imaging plane and wavelength of $\lambda=488 nm$. }
Using samples with
homogeneous fluorescence we calibrated for the non-uniform intensity
of the TIRF excitation field in the image plane. In order to analyze
the data, we classified all filament segments closer than 80 nm to the
cover slide as adsorbed.  To investigate the influence of chamber
thickness, we have studied the adsorption/desorption of actin
filaments in chambers with thickness 2 and 10 $\mu$m. Such chambers
were prepared by using 2 and 10 $\mu$m beads as spacers.

\section{Simulations}
The adsorption experiments have been complemented by off-lattice Monte
Carlo computer simulations of a single non-grafted polymer filament of
length $L$ and persistence length $L_p$, confined between two
flat walls separated by a distance $L_z$.  For not too long polymers
($L/L_p\approx 1$), in which self-avoidance effects are negligible, a
description in the worm-like chain model is
appropriate~\cite{HSU010,HSU011}.  Each semi-flexible polymer was
described by $N$ inextensible infinitely thin elements of length
$l_0=L/N$ and a bending potential $V_b=\kappa(1-{\bf t}_i \cdot {\bf
  t}_{i+1})/l_0$ between subsequent segments. Here ${\bf t}_i$ is the
unit tangent vector at the $i$-th segment in the chain, and $\kappa$
the bending stiffness. The persistence length $L_p$ is defined by
$\av{\vc{t}_i\cdot\vc{t}_j}/d=\exp{[-(d-1)s_{ij}/L_p]}$ for unconfined
filaments, where $s_{ij}$ is the distance between two segments along
the contour and $d$ the embedding dimensionality. The persistence
length is related to the bending rigidity by $L_p=2\kappa/k_BT$ with
$k_BT$ the thermal energy~\cite{DOI,KIERFELD03}. The filament
interaction with the walls is included as a hard-core repulsive
interaction. Similar to the experimental system, one of the walls (in fact the lower one) is
made attractive by superimposing an attractive square well interaction
of range $\delta$ and depth $\varepsilon=\widetilde{\varepsilon}/k_BT$
on top of the repulsive interaction. Here $\widetilde{\varepsilon}$ is
the bare interaction strength per unit length of polymer.  
The interaction strength was varied by changing $\widetilde{\varepsilon}$
and keeping the temperature (and thus the persistence length)
constant. This is similar to depletion induced attractions where the
strength of the attraction is tuned by the concentration of depletant
and not by variation of temperature.

The Monte Carlo simulations involved pivot configurational moves in
two and three dimensions~\cite{LAL69,MADRAS88}, uniform translations
perpendicular to the wall and uniform rotational moves. Parameters in
the simulations were chosen to resemble those in the experiments, and
results are discussed in terms of dimensionless quantities. We used
filaments of $L_p=16$, and $L=33$, such that $L/L_p \simeq 2.1$.  For
a large enough number of segments $N$ ($l_0$ small enough) the results
are independent of the discretization. $N$ was estimated from the
transversal deviation $\av{b^2(s)}=2s^3/3L_p$ of a semi-flexible
polymer from its initial direction~\cite{ODIJK83}. By identification
of $s$ with the segment length $L/N$, $b$ with the range of the
attraction and arguing that the transversal crossing of the attraction
well should require more than one step we find that $s <
(3L_p\delta^2/2)^{1/3}$.  Averages were calculated over adsorbed
configurations which have non-zero filament fraction within a cut-off
distance $R_c$ from the wall. The parameters chosen for the attractive
wall are $\delta=0.01$ such that $\delta/L \simeq 0.0004$ and
$R_c=8\,\delta$. Our work examined the effect of confinement on
adsorption transition of filament with both free ends, in contrast to
other simulations that examined transitions of unconfined filaments
which were grafted at one end~\cite{HSU013,HSU013slit}.  
We calculate the average length of tails, trains and loops and
normalize them by the filament persistence length $L_p$, since it is
uniquely defined in both simulations and experiments. This in contrast
to the polymer contour length $L$ which is subject to the intrinsic
polydispersity of experimental samples.

\section{Results \& discussion}
At low depletant concentrations (weak adsorption limit) the vast
majority of filaments are largely desorbed from the wall
(Fig.\ref{fig:definitions}b,c-i). When such samples are viewed with
TIRF microscopy, one occasionally observes short filament segments
that briefly remain in the evanescent field. Increasing the depletant
concentration leads to a transition regime where the length of the
adsorbed segments increases (Fig.\ref{fig:definitions}b,c-ii). Here,
filaments remain attached to the surface on timescales longer than the
duration of the experiment while exhibit large fluctuations. At even
higher depletant concentrations (strong adsorption limit) almost
entire filament is permanently adsorbed on the interface. One can
occasionally identify desorbed short segments that are either located
in the middle of the filament or at its
ends. (Fig.\ref{fig:definitions}b,c-iii, for related movies see
Ref.~\cite{movie}). 
Qualitatively similar filament behavior is observed in computer
simulations, Fig.~\ref{fig:definitions}d shows typical snapshots for
adsorption strengths where the filament is weakly bound (i), where a
loop is present (ii) and for strong attractions where the filament is
tightly bound to the surface and exhibits only one train (iii).

\begin{figure}[h]
\includegraphics[width=0.46\textwidth]{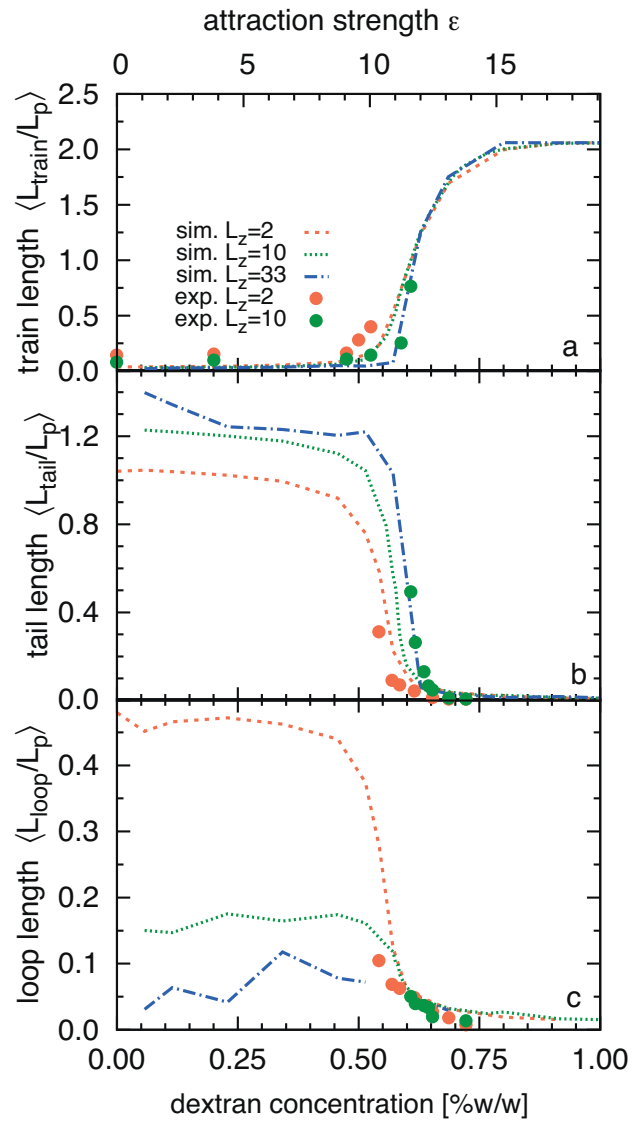}
\caption{The length of filament trains, tails and loops and their
  dependence on the attraction strength, $\varepsilon$ (or equivalent
  depletant concentration $C_p$), averaged over independent
  configurations and normalized by the filament persistence length
  $L_p$. \textbf{a)}~Average train length
  $\av{L_{\mbox{\scriptsize{train}}}}/L_p$.  \textbf{b)}~Average tail
  length $\av{L_{\mbox{\scriptsize{tail}}}}/L_p$. \textbf{c)}~Average
  loop length $\av{L_{\mbox{\scriptsize{loop}}}}/L_p$. Filled circles
  correspond to experimental data and lines indicate simulation
  results. Chamber heights, $L_z$ are specified in the figure
  legend. The simulation date are scaled onto depletant concentration,
  $C_p$, using $\varepsilon=C_p/C$ were $C=16.75$. }
\label{fig:transitionsL} 
\end{figure}

  To quantify these observations we have measured how the average
  length of filament trains, tails and loops depend on the strength of
  the filament-wall attraction. For strongest attractions,
  fluctuations away from the surface are almost completely suppressed
  and the average train length $\av{L_{\mbox{\scriptsize{train}}}}$
  approaches the filament contour length, $L$
  (Fig.\ref{fig:transitionsL}a). In this limit the average length of
  tails and loops is negligible. With decreasing attraction strength
  the filament starts to fluctuate away from the minimum energy state
  by locally desorbing from a flat surface, at first slightly but in
  the transition regime $\langle L_{\mbox{\scriptsize{train}}}
  \rangle$ drops precipitously, approaching almost zero for very weak
  attractions (desorbed filaments). The transition regime is clearly
  indicative of the location of the filament adsorption-desorption
  transition. However, we note that for finite sized filaments, as
  those under consideration here, the adsorption transition is not a
  sharply defined singular point but rather occurs over a finite range
  of relevant parameters~\cite{MOEDDEL011,HSU013}, and can slightly
  vary when considering loops, trains or tails.  A decrease in
  $\langle L_{\mbox{\scriptsize{train}}} \rangle$ means that a large
  fraction of the filament is desorbed. Consequently, it is
  accompanied with a simultaneous sharp increase in average loop,
  $\langle L_{\mbox{\scriptsize{loop}}} \rangle$, and tail length
  $\langle L_{\mbox{\scriptsize{tail}}} \rangle$. As can be observed
  in Fig.\ref{fig:transitionsL} the adsorption-desorption transition
  can be identified by either a rapid increase in $\langle
  L_{\mbox{\scriptsize{tail}}} \rangle$ or a rapid decrease in
  $\langle L_{\mbox{\scriptsize{train}}} \rangle$.

Next we compare experimental data to simulations performed using
equivalent molecular parameters. To achieve this the simulation
attraction strength, $\varepsilon$, is scaled with a fitting factor
$C$, defined as $\varepsilon=C_p/C$, where $C_p$ is the depletant
concentration. The best collapse of simulation data to experiments
determines the magnitude of $C$. We use the same value of $C$ for all
our subsequent analyses. In principle, one should be able to compute
$C$ from relevant experimental parameters such as the filament surface
charge and charge of the glass substrate, the depletant polydispersity
and the precise shape of the attraction potential. However, since some
of these parameters are difficult to estimate for now we simply use
$C$ as a scaling parameter. $C=16.75$ yields the best quantitative
agreement with the experimental measurements over the entire
measurement range (Fig.~\ref{fig:transitionsL}).

\begin{figure}[h]
\includegraphics[width=0.46\textwidth]{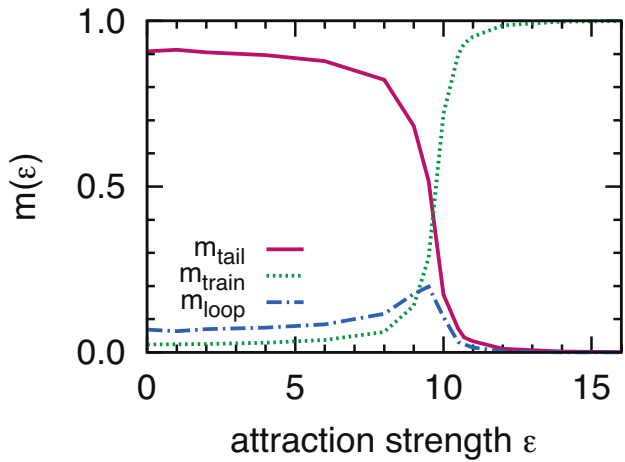}
\caption{\label{fig:fraction1} Fraction of the total filament length
  that is stored in tails, loops and trains as a function of
  filament-wall attraction strength $\varepsilon$.  The simulations
  were performed for $L_z=2$.}
\end{figure}

In the desorbed and transition regimes the average tail length is
considerably larger than the average loop length
$\av{L_{\mbox{\scriptsize{tail}}}}>\av{L_{\mbox{\scriptsize{loop}}}}$
indicating the dominance of tail over loop fluctuations
(Fig.~\ref{fig:transitionsL}b,c). %
Filament ends are linked to the adsorbed filament segments only on one
side, thus they exhibit more pronounced fluctuations and have higher
entropy, when compared to loops which are linked to adsorbed segments
at both sides. Consequently, for finite sized filaments, the desorbed
segments are more likely to be located at the filament ends. To depict
this effect clearly, we plot $m_{train}$, $m_{loop}$, and $m_{tail}$,
the average filament length stored in trains, loops, and tails
normalized with $L$ (Fig.~\ref{fig:fraction1}). These fractions
add up to unity. Data plotted in this way confirms that when compared
to tails, loops do not play an important role in the desorption of
semi-flexible filaments. This is in contrast to longer flexible
polymers, where loops dominate the desorption process~\cite{HSU013};
and also in contrast to stiff rod-like polymers, where loops cannot
exist.

\begin{figure}[h]
\includegraphics[width=0.46\textwidth]{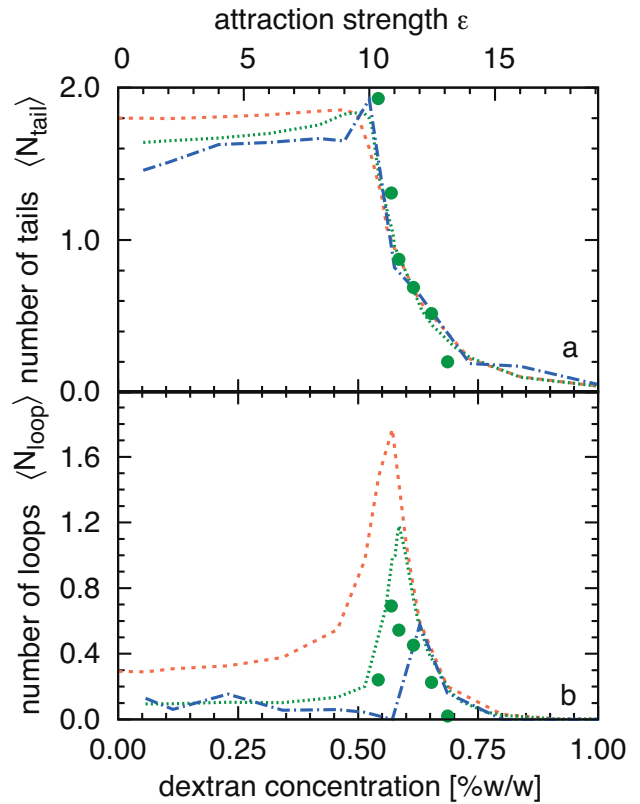}
\caption{\label{fig:transitionsN} \textbf{a}) Average number of tails
  $\av{N_{\mbox{\scriptsize{tail}}}}$ plotted as a function of the
  attraction strength, $\varepsilon$ (or equivalent the depletant concentration $C_p$).  \textbf{b}) Average number of
  loops $\av{N_{\mbox{\scriptsize{loop}}}}$ as a function of
  $\varepsilon$.  Symbols, lines and color coding is as in
  Fig.\protect{\ref{fig:transitionsL}}}
\end{figure}

It is also possible to elucidate the nature and location of the
adsorption transition by examining how the average number of loops
$\av{N_{\mbox{\scriptsize{loop}}}}$ and tails
$\av{N_{\mbox{\scriptsize{tail}}}}$ changes with the filament-wall
attractions (Fig.~\ref{fig:transitionsN}). In the strongly adsorbed
regime the train length approaches the filament contour length; there
is on average only one train and there are very few short loops and
tails. Decreasing polymer concentration increases the frequency of
configurations with tails and loops. The transition regime can be
characterized by a sharp increase in the number of loops and
trains. At the transition point $\av{N_{\mbox{\scriptsize{tail}}}} \sim 2$
indicating that essentially all filament configurations have both ends
desorbed. Furthermore,
$\av{N_{\mbox{\scriptsize{tail}}}}>\av{N_{\mbox{\scriptsize{loop}}}}$, indicates
that tails dominate over loops, in agreement with previous
findings. Finally, below the adsorption transition loops merge with
other loops and tails, so that the number of loops decreases and the
length of the tails increases.  Quantitative agreement between
numerical simulations and experiments is obtained using the same
scaling parameter, $C$, used in Fig.~\ref{fig:transitionsL}.

Increasing polymer confinement decreases the configurational space
available to desorbed filaments and thus shifts the
adsorption-desorption transition. To explore this effect we have
repeated experiments for chambers with two different thicknesses
(Fig.~\ref{fig:transitionsL}). For strongly bound filaments tails and
loops do not extend very far in the bulk and thus cannot interact with
the opposite surface. In this regime all measured quantities are
independent of confinement. For intermediate attraction strengths,
confinement suppresses configurations with tails that extend far into
the bulk. Since long tails appear in a configuration either alone or
in combination with small loops, removal of these configurations
necessarily leads to an effective increase of the average loop
size. For weakly bound filaments, the average number of loops
decreases strongly but their average size saturates to a constant
value which becomes larger with increasing confinement. Changing the
number of accessible configurations of the desorbed filament changes
the balance between the entropy of the desorbed state and the energy
of the adsorbed state, effectively shifting the location of the
adsorption/desorption transition to lower attraction strengths.  This
trend is clear in Fig.~\ref{fig:transitionsL} where experiments with
chambers of different height are shown to be in reasonable agreement
with simulations performed with different wall separations.

\begin{figure}
\includegraphics[width=0.45\textwidth]{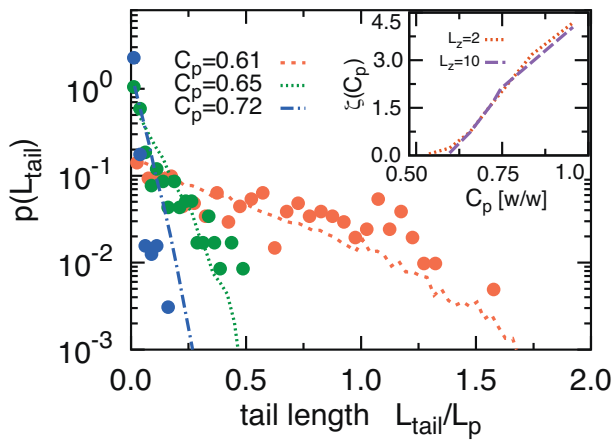}
\caption{Probability distribution of tail lengths
  p$(L_{\mbox{\scriptsize{tail}}})$ at three different attraction
  strengths (depletant concentrations). Filled circles indicate
  experimental data taken for chambers with $L_z=10$ while lines
  indicate simulations results.
  Inset:
  Variation of the inverse decay length $\zeta$ for simulations
  with $L_z=2$ and $L_z=10$.}\label{fig:distr}
\end{figure}

Given the dominant influence of the tails on the adsorption-desorption
transition we investigated in more depth the distribution of tail
lengths, $p(L_{\mbox{\scriptsize{tail}}})$ (Fig.\ref{fig:distr}). In
the strong attraction limit the filaments are almost completely
adsorbed. Therefore the average tail length is small which leads to a
very fast decay of $p(L_{\mbox{\scriptsize{tail}}})$. This
distribution can be described with the exponential function
\begin{equation}
p(L_{\mbox{\scriptsize{tail}}}) \sim \exp[-L_{\mbox{\scriptsize{tail}}} \zeta(C_p)], 
\end{equation}
where the factor $\zeta(C_p)$ has units of an inverse length and
increases with increasing attraction strength. Upon approaching the
adsorption transition, the distribution becomes rather flat and
$\zeta(C_p)$ eventually vanishes, showing that all tail lengths are
equally probable.  In this regime, $p(L_{\mbox{\scriptsize{tail}}})$
can no longer be described by an exponential decay. Once the tail
length becomes of the order of the filament contour length, the
filament completely desorbs from the surface and
$p(L_{\mbox{\scriptsize{tail}}})$ shows a maximum around
$L_{\mbox{\scriptsize{tail}}}=L$. Using the same scaling parameter,
$C$, discussed previously we can quantitively fit the simulation data
to experimentally measured probability histograms of tail lengths. 
The inset of Fig.~\ref{fig:distr} shows that at strong attractions
$p( L_{\mbox{\scriptsize{tail}}})$ does not depend on
confinement~\cite{note:hsu-expon}. In this limit the tails are rather
close to the lower surface so that they do not feel the confinement
effects.

\begin{figure}
\includegraphics[width=0.45\textwidth]{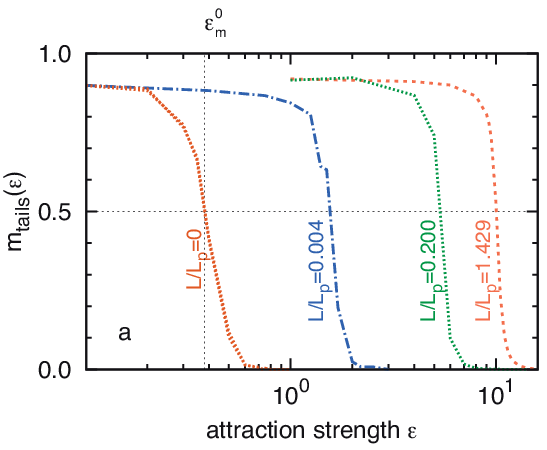}
\includegraphics[width=0.45\textwidth]{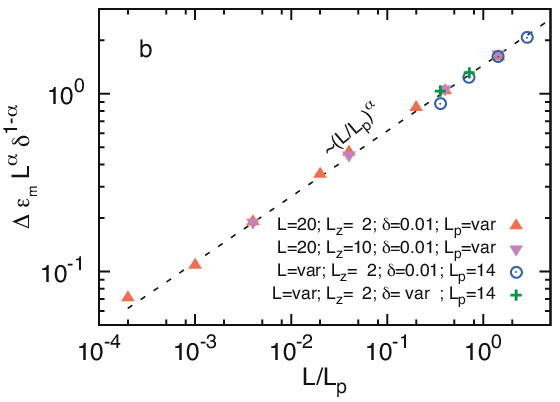}
\caption{\textbf{a}) Fraction of the total filament length stored in
  tails $m_{\mbox{\scriptsize{tail}}}(\varepsilon)$ as a function of
  the attraction strength $\varepsilon$. Simulations with $L=20$ and
  $L_z=10$ for various persistence lengths. The attraction strength
  $\varepsilon_{m}^0$ for which half of the infinitely stiff filament
  is in tails is explicitly indicated. \textbf{b}) Scaled variation
  $\Delta \varepsilon_{m}$ as a faction of $L/L_p$ with
  $\alpha=0.37$. Symbols correspond to simulations with four sets of
  parameters indicated in the labels (crosses correspond to
  $(\delta,L)=(0.005,5)$ and $(0.0025,10)$). The dashed line is a fit to a
  power law dependence $(L/L_p)^\alpha$ .}\label{fig:fraction}
\end{figure}

Semi-flexible filaments have more degrees of freedom when compared to
rigid filaments. Consequently, confining such filaments onto a wall
incurs a larger entropic cost which shifts the adsorption transition
to stronger attractions. We investigate this effect by plotting the
average fraction of filament length stored in the tails
$m_{\mbox{\scriptsize{tail}}}(\varepsilon)=
\av{N_{\mbox{\scriptsize{tail}}}(\varepsilon)}
\av{L_{\mbox{\scriptsize{tail}}}(\varepsilon)}/L$ for filaments with
varying stiffness. In the strongly adsorbed limit
$m_{\mbox{\scriptsize{tail}}}(\varepsilon)$ vanishes while in the
opposite weak adsorption limit the same quantity approaches a constant value that depends on $\delta/L$
(Fig.\ref{fig:fraction}a). For decreasing $\delta/L$ this plateau value approaches unity. Following previous work~\cite{HSU013}, the
location of the adsorption transition is identified as the attraction
value $\varepsilon_m$ at which half the filament length is desorbed
$m_{\mbox{\scriptsize{tail}}}(\varepsilon_m)=0.5$. Increasing the
filament flexibility shifts the adsorption transition to larger
$\varepsilon_m$ (Fig.~\ref{fig:fraction}a). For example, the
attraction strength necessary to adsorb a rigid filament
$\varepsilon_m^0$ is approximately one order of magnitude lower than
the attraction necessary for adsorption of a filament whose
persistence length is equal to its contour length. The effect of
flexibility on the critical adsorption energy $\varepsilon_m$ can be
characterized by the displacement of the attraction at which the
adsorption transition occurs as
$\Delta\varepsilon_m=\varepsilon_m-\varepsilon_m^0$.  Plotting
$\Delta\varepsilon_m$ as a function of $L/L_p$ reveals a scaling
relationship $\Delta\varepsilon_m \sim (L/L_p)^\alpha$ where
$\alpha=0.37$ (Fig.~\ref{fig:fraction}b). Simulations for different
thickness chambers yields the same power law increase, indicating that
$\Delta\varepsilon_m$ is independent of confinement. Since
$\varepsilon$ has units of inverse length, a dimensionless quantity
$\Delta\varepsilon_mL^\beta\delta^{1-\beta}$can be constructed using
the remaining relevant length scales of the problem. The dependency of
$\Delta \varepsilon_m$ on the relevant length scales is further tested
by varying $L$ at fixed $L_p$, varying $L/L_p$ at fixed $\delta$, and
simultaneously varying $L$ and $\delta$ at fixed $L_p$. The resulting
four data sets collapse onto a single line by using the fitted
exponent $\beta=0.37$ (Fig.~\ref{fig:fraction}b). Equivalence of
$\alpha$ and $\beta$ implies that $\Delta\varepsilon_m$ is independent
of the filament length. The final result for our scaling relationship
is is summarized as
\begin{equation}\label{adsorp.scale}
\Delta \varepsilon_m L_p^\alpha \delta^{1-\alpha} \sim 1.
\end{equation}
Our model thus can predict the adsorption of a semi-flexible polymer
$\varepsilon_m$ if the adsorption transition of a corresponding stiff
rod ($\varepsilon^0_m$) is known. For the simple square well potential
used in the simulations, for small values of $\delta$ and for $L_z\leq
L+\delta$, $\varepsilon_m^0$ can be obtained analytically as
\begin{equation} \label{epsm.rod}
  \varepsilon_m^0=\frac{1}{L}\ln\left(\frac{2L_z}{\delta}\right).
\end{equation}
Details of the calculation outlined above and related discussion will
be presented elsewhere.

The origin of scaling in Figure~\ref{fig:fraction}b remains
unclear. However, it is intriguing that a similar combination of
lengths arises in derivation of the Odijk's deflection length, $s$,
which characterizes the fluctuations of filaments confined to a
narrow tube~\cite{ODIJK83,ODIJK08}. When the transverse deviation
of confined filaments is of the order of the filament spacing $D$,
Odijk's deflection length is defined as $s\sim
L_p^{\alpha}D^{1-\alpha}$ with $\alpha=1/3$.  If we assume that for
strongly confined polymers the attraction range of an adsorbed
polymer, $\delta$ is equivalent to $D$ we find scaling $\sim
L_p^{1/3}\delta^{2/3}$ which is close to, but not identical to
Eq.~\ref{adsorp.scale}. This result suggests that the fluctuations
 of adsorbed filaments might be analogues to fluctuations of
filaments that are confined by an effective tube.

\section{Conclusions}
In summary, we have combined TIRF microscopy, depletion interactions,
semi-flexible F-actin filaments and computer simulations to study the
adsorption/desorption of semi-flexible polymers. Real-space
visualization enabled us to extract important yet difficult to measure
statistics of trains, tails and loops of filaments undergoing the
adsorption transition. Our work visually demonstrates the importance
of the internal degrees of freedom when considering adsorption of
polymers onto a surface.  Computer simulations quantitatively agree with
  experimental results and have enabled us to gain detailed
insight in how decreasing the polymer stiffness from a infinitely
stiff rod shifts the adsorption transition to larger attraction
values. Furthermore, the description of the data in terms of a
universal power law dependence opens up the promising possibility of
determining the location of the adsorption transition in terms of the
polymer length, persistence length, wall separation and attraction
range.  The similarity of the scaling law with and established theory
of polymer confinement in the absence of adsorption is an interesting
indication that adsorption and confinement share similar physical
grounds. Our theoretical predictions could be further tested
in experiments in which the filament persistence length is tuned while
keeping all other parameters constant. This might be possible by using
filamentous phages, where a point
mutations of the major coat protein can tune the filament flexibility
by up to 400\%~\cite{BARRY}. The range of the attraction can be
modified by using a different size of depleting polymer. From a
practical perspective, the surface-induced depletion interactions
studied here have been used to assemble diverse soft materials such as
extensile microtubule based 2D active nematics, supercoiling actin rings and
contractile actin gels~\cite{SANCHEZ,MURELL,KULIC}. Similar fluctuations are
also important when instead of a surface a filament binds onto another
filament.  For example, only after accounting for such fluctuations
could the effective measured strength of the depletion interaction
between a pair of filaments be fitted to a theoretical
model~\cite{LAU}.

\begin{acknowledgments}
  We would like to acknowledge Aggeliki Tsigri, Donald Guu and Dzina
  Kleshchanok for their collaboration in an initial phase of this
  work. We thank Jan Kierfeld and Hsiao-Ping Hsu for valuable
  discussions. DW and ZD acknowledge support of NSF-CMMI-1068566 and
  NSF-CAREER-0955776. We also acknowledge the use of MRSEC optical
  microscopy facility which is supported by grants NSF-MRSEC-0820492
  and NSF-MRI-0923057.
\end{acknowledgments}


\begin{thebibliography}{9}
\bibitem{note:hsu-expon} Hsu and Binder have characterized the
distribution function of loops for grafted long semi-flexible polymers
finding a behavior consistent with a power law decay (see Fig.~20
in~\cite{HSU013slit}).
\end{thebibliography}
\end{document}